\documentclass[twocolumn,english,aps,pra]{revtex4-2}
\usepackage[T1]{fontenc}
\usepackage{amsmath,graphicx,amssymb,epsfig,babel,dsfont}
\usepackage{mathrsfs}
\usepackage{xcolor}
\usepackage[normalem]{ulem}

\renewcommand{\Re}{{\rm Re}}
\renewcommand{\Im}{{\rm Im}}
\newcommand{\Tr}{{\rm Tr}}
\newcommand{\rd}{{\rm d}}

\newcommand{\ri}{{\rm i}}

\newcommand{\re}{{\rm e}}

\newcommand{\kB}{k_{\rm B}}

\begin{document}

\title{On persistent energy currents at equilibrium in non-reciprocal systems}

\author{S.-A. Biehs}
\email{s.age.biehs@uni-oldenburg.de}
\affiliation{Institut f\"{u}r Physik, Carl von Ossietzky Universit\"{a}t, 26111, Oldenburg, Germany}

\author{Ivan Latella}
\email{ilatella@ub.edu}
\affiliation{Departament de F\'{i}sica de la Mat\`{e}ria Condensada, Universitat de Barcelona, Mart\'{i} i Franqu\`{e}s 1, 08028 Barcelona, Spain}
\affiliation{Institut de Nanociència i Nanotecnologia de la Universitat de Barcelona (IN2UB), Diagonal 645, 08028 Barcelona, Spain}

%
%


\begin{abstract}
		We investigate the properties of the mean Poynting vector in global thermal equilibrium, which can be non-zero in non-reciprocal electromagnetic systems. Using dyadic Green's functions and the fluctuation-dissipation theorem, we provide a general proof that the mean Poynting vector is divergence-free under equilibrium conditions. Relying on this proof, we explicitly demonstrate that for systems where a normal mode expansion of the Green's function is applicable, the divergence of the equilibrium mean Poynting vector vanishes. As concrete examples, we also examine the equilibrium mean Poynting vector near a planar non-reciprocal substrate and in configurations involving an arbitrary number of dipolar non-reciprocal objects in free space. Finally, we argue that the so-called persistent heat current, while present in equilibrium, cannot be detected through out-of-equilibrium heat transfer measurements.
\end{abstract}

\maketitle

%
%

\section{Introduction}

\begin{figure}
	\includegraphics[width=0.45\textwidth]{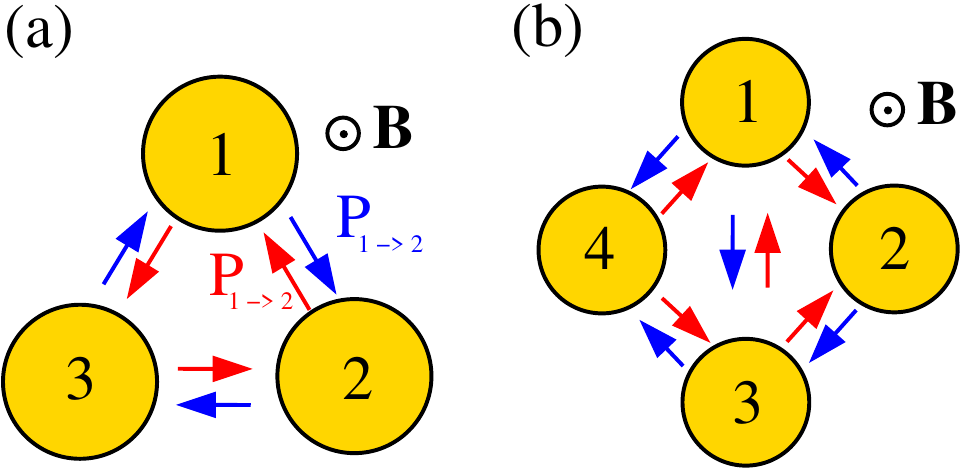}
	\caption{(a) Persistent heat current: The spherical nanoparticles $i = 1,2,3$ are in thermal equilibrium with the environment. A transferred power $P_{i \rightarrow j}$ from particle $i$ to particle $j$ can be defined even in equilibrium. When a magnetic field is applied one finds $P_{i \rightarrow j} \neq P_{j \rightarrow i}$, but since for the heat current in clockwise direction we have $P_{1 \rightarrow 2} = P_{2 \rightarrow 3} =  P_{3 \rightarrow 1} $ and for the anti-clockwise direction $P_{1 \rightarrow 3} = P_{3 \rightarrow 2} =  P_{2 \rightarrow 1}$, there is no net heat transfer~\cite{ZhuFan}. (b) Hall effect for thermal radiation: The temperatures $T_4 > T_2$ are fixed. In steady state, due to the exchanged heat by thermal radiation one finds $T_1 \neq T_3$ when a magnetic field is applied. Again, this is due to the fact that $P_{i \rightarrow j} \neq P_{j \rightarrow i}$ in that case~\cite{PBAHall}.
		\label{Sketch} }
\end{figure}

About one decade ago Zhu and Fan~\cite{ZhuFan} considered three magneto-optical nanoparticles in a $C_3$ symmetric configuration in global thermal equilibrium, as shown in Fig.~\ref{Sketch}(a). They showed that the nanoparticles exchange heat even when the system is in equilibrium, resulting in an apparent heat flow, either clockwise or anti-clockwise, without any net energy transfer to the particles.
This can be understood in the following way. The transferred power $P_{i \rightarrow j}$ and $P_{j \rightarrow i}$ ($i,j = 1,2,3$) from particles $i$ to particle $j$ and vice versa are in general equal. However, by applying an external magnetic field the time-reversal symmetry breaking results in $P_{i \rightarrow j} \neq P_{j \rightarrow i}$, so that a constant circular current in clockwise or anti-clockwise direction arises with the property that  $P_{1 \rightarrow 2} = P_{2 \rightarrow 3} =  P_{3 \rightarrow 1} $ and $P_{1 \rightarrow 3} = P_{3 \rightarrow 2} =  P_{2 \rightarrow 1}$, in such as way that there is no net heat transfer to any particle. This heat current is called {\itshape persistent heat current} and it exists in any many-body configuration with more than two non-reciprocal objects~\cite{ZhuEtAl2018}. 
The persistent heat flux, spin, and angular momentum connected to the persistent heat current, however, also exists in the vicinity of single non-reciprocal objects and has been studied for planar interfaces~\cite{KhandekarSpin}, single nanoparticles~\cite{OTTcircular}, and planar cavities~\cite{Silveirinha}. As argued by Zhu and Fan~\cite{ZhuFan}, the existence of this persistent heat current does not violate the second law of thermodynamics, since the total net heat flow into each particle vanishes when all the objects have the same temperature.

In the same year of the work on the persistent heat current~\cite{ZhuFan}, Ben-Abdallah~\cite{PBAHall} considered four magneto-optical nanoparticles in a $C_4$ symmetric configuration in an out-of-equilibrium situation, as shown in Fig.~\ref{Sketch}(b). As in the previous three-particle configuration, a persistent heat current would naturally exist if the system is considered in thermal equilibrium. For the non-equilibrium situation, Ben-Abdallah showed that a steady-state temperature difference is established between the particles at the top and bottom when a temperature difference is applied between the particles on the left and right as well as an external magnetic field. This effect is called {\itshape Hall effect for thermal radiation} or {\itshape photon thermal Hall effect} and has been further investigated in Refs.~\cite{OTTReview,Cuevas}. As for the persistent heat current, this phenomenon is a consequence of the time-reversal symmetry breaking introduced by the magnetic field, making the system non-reciprocal and leading to transferred powers satisfying $P_{i \rightarrow j} \neq P_{j \rightarrow i}$. Consequently, the persistent heat current and the Hall effect for thermal radiation both exist also for other non-reciprocal materials such as Weyl semi-metals~\cite{OTTPBAHall}.

Following these discoveries, the existence of a persistent heat current has motivated researchers to propose systems in which it might be experimentally observed. One such suggestion involves placing atoms or molecules in close proximity to a planar interface and measuring the momentum transfer induced by the persistent heat flux near the interface~\cite{KhandekarSpin}. However, such a net momentum transfer would violate the second law of thermodynamics, and it has been explicitly demonstrated that no momentum transfer can occur under thermal equilibrium conditions~\cite{Krueger}. Another study~\cite{GuoEtAl2019} proposes that applying a small temperature difference could enable the measurement of the persistent heat current, arguing that it is formally related to the photon thermal Hall effect and that both phenomena essentially arise from the same underlying mechanism.

In this work, we revisit the relationship between persistent heat flux and heat transfer in non-reciprocal systems. We present a general proof that the mean Poynting vector is divergence-free in thermal equilibrium. This result is further supported by explicit verifications in representative cases such as the region near a non-reciprocal planar substrate and in an arbitrary assembly of non-reciprocal nanoparticles. Finally, we discuss why the persistent heat current cannot be accessed through out-of-equilibrium heat transfer experiments and clarify the fundamental differences between this phenomenon and the Hall effect for thermal radiation.
%
%
%

\section{Fluctuation dissipation theorem and Poynting vector}

Let us consider the electric and magnetic fields $\mathbf{E} (\mathbf{r}, t)$ and $\mathbf{H} (\mathbf{r}', t')$ at different positions $\mathbf{r}$ and $\mathbf{r}'$ and times $t$ and $t'$, respectively, in a system in global thermal equilibrium at temperature $T$. The correlation function involving the Cartesian components $E_\alpha (\mathbf{r}, t)$ and $H_\beta(\mathbf{r}', t')$ of these fields
can be obtained within the framework of the linear response theory and is given by~\cite{GSA}
\begin{equation}
	\langle E_\alpha (\mathbf{r}, t) H_\beta(\mathbf{r}', t') \rangle_{\rm eq}^{\rm sym} = \int_{-\infty}^{\infty} \!\! \frac{\rd \omega}{2 \pi} \psi^{\rm EH}_{\alpha,\beta} (\mathbf{r}, \mathbf{r'};\omega) \re^{- \ri \omega(t - t')}
\end{equation}
with the spectral function
\begin{equation}
\begin{split}
	\psi^{\rm EH}_{\alpha\beta} (\mathbf{r}, \mathbf{r'};\omega) &= \Theta(\omega,T) 2 \omega \mu_0 \\ 
	& \times \biggl(-\frac{\ri}{2}\biggr)  \biggl[ \mathds{G}^{\rm EH}_{\alpha\beta}(\mathbf{r},\mathbf{r}';\omega) - {\mathds{G}^{\rm HE}_{\beta\alpha}}^*(\mathbf{r}',\mathbf{r};\omega) \biggr],
\end{split}
	\label{Eq:Psispec}
\end{equation}
where
\begin{equation}
   \Theta(\omega,T) = \frac{\hbar \omega}{\re^{\hbar\omega/\kB T} - 1} + \frac{\hbar\omega}{2}
\end{equation}
is the mean energy of a harmonic oscillator at frequency $\omega$, $\mu_0$, $\hbar$, and $\kB$ being the vacuum permeability, the reduced Planck constant, and Boltzmann constant, respectively.
In Eq.~(\ref{Eq:Psispec}), we have introduced the dyadic Green's functions $\mathds{G}^{\rm EH}$ and $\mathds{G}^{\rm HE}$ which are the electric field response to a magnetic dipole and the magnetic field response to an electric dipole~\cite{Eckhardt}, respectively. 

Before we proceed, we want to point out that the correlation function in Eq.~(\ref{Eq:Psispec}), also termed the fluctuation-dissipation theorem of the first kind~\cite{Eckhardt}, is a quantum mechanical expression for the expectation value of the electric and magnetic field operators $\hat{E}_\alpha (\mathbf{r}, t)$ and $\hat{H}_\beta(\mathbf{r}', t')$ in symmetrical order, which means that
\begin{equation}
	\langle E_\alpha (\mathbf{r}, t) H_\beta(\mathbf{r}', t') \rangle_{\rm eq}^{\rm sym} = \frac{1}{2}\bigl\langle \bigl\{\hat{E}_\alpha (\mathbf{r}, t), \hat{H}_\beta(\mathbf{r}', t') \bigr\} \bigr\rangle_{\rm eq} ,
\end{equation}
where the bracket $\{\circ,\circ \}$ stands for the anti-commutator. This correlation function of the symmetrically ordered operators is related to the corresponding correlation functions for the normally and anti-normally ordered operators~\cite{GSA}. Within the framework of fluctuational electrodynamics, the fields are typically treated as classical quantities and the quantum expressions for the expectation value of the symmetrically ordered operators are employed for the computation of their correlation functions.

In the following, we are only interested in the equal time correlation function. It is straightforward to show that in this case
\begin{equation}
	\langle E_\alpha (\mathbf{r}, t) H_\beta(\mathbf{r}', t) \rangle_{\rm eq}^{\rm sym} = \int_{0}^{\infty} \!\! \frac{\rd \omega}{2 \pi} 2 \Re \bigl[ \psi^{\rm EH}_{\alpha\beta} (\mathbf{r}, \mathbf{r'};\omega) \bigr],
	\label{Eq:Equaltimecorrelation}
\end{equation}
where we used that $\psi^{\rm EH}_{\alpha\beta} (\mathbf{r}, \mathbf{r'};-\omega) = {\psi^{\rm EH}_{\alpha\beta}}^* (\mathbf{r}, \mathbf{r'};\omega)$ which follows from the symmetry relation $\Theta(-\omega,T) = \Theta(\omega,T)$ as well as from $\mathds{G}^{\rm EH}(\mathbf{r},\mathbf{r}';- \omega) = {\mathds{G}^{\rm EH}}^*(\mathbf{r},\mathbf{r}'; \omega)$ and $\mathds{G}^{\rm HE}(\mathbf{r},\mathbf{r}';- \omega) = {\mathds{G}^{\rm HE}}^*(\mathbf{r},\mathbf{r}'; \omega)$. Furthermore, for non-magnetic materials, the dyadic Green's function $\mathds{G}^{\rm EE}$ describing the electric field response due to an electric dipole is connected with the mixed Green's functions by the relations~\cite{Chew}
\begin{align}
	\mathds{G}^{\rm EH}(\mathbf{r,r'};\omega) &= \frac{1}{\ri \omega \mu_0} \mathds{G}^{\rm EE}(\mathbf{r,r'};\omega)\times\nabla',  \label{Eq:GEHEE}  \\
	\mathds{G}^{\rm HE}(\mathbf{r,r'};\omega) &= \frac{1}{\ri \omega \mu_0} \nabla\times\mathds{G}^{\rm EE}(\mathbf{r,r'};\omega), \label{Eq:GHEEE} 
\end{align}
where $\nabla=\partial/\partial\mathbf{r}=(\partial_x,\partial_y,\partial_z)$ and in the same way $\nabla'=\partial/\partial\mathbf{r}'=(\partial'_x,\partial'_y,\partial'_z)$.
Note that, here and throughout the following, any differential operator appearing to the right of the Green’s tensor is understood to act to the left. The above expressions (\ref{Eq:GEHEE}) and (\ref{Eq:GHEEE}) can be generalized by replacing the permeability of vacuum $\mu_0$ by the corresponding inverse permeability tensor of the medium~\cite{Chew}.

Throughout the paper, we focus on the mean Poynting vector in thermal equilibrium, defined in the standard manner by 
\begin{equation}
\begin{split}
	\langle \mathbf{S} \rangle_{\rm eq} &=  \langle \mathbf{E} (\mathbf{r}, t) \times \mathbf{H}(\mathbf{r}, t) \rangle_{\rm eq}^{\rm sym} \\
		  &= \lim_{\mathbf{r}' \rightarrow \mathbf{r}} \langle \mathbf{E} (\mathbf{r}, t) \times \mathbf{H}(\mathbf{r}', t) \rangle_{\rm eq}^{\rm sym} 
\end{split}
	\label{Eq:Pontingvector}
\end{equation}
or, in components, by
\begin{equation}
\begin{split}
	\langle S_\nu \rangle_{\rm eq} &= \epsilon_{\nu \alpha \beta} \langle E_\alpha (\mathbf{r}, t) H_\beta(\mathbf{r}, t) \rangle_{\rm eq}^{\rm sym} \\
	&= \lim_{\mathbf{r}' \rightarrow \mathbf{r}}\epsilon_{\nu \alpha \beta} \langle E_\alpha (\mathbf{r}, t) H_\beta(\mathbf{r}', t) \rangle_{\rm eq}^{\rm sym}
\end{split}
	\label{Eq:Pontingvector_components}
\end{equation}
with the summation convention for repeated Greek (Cartesian) indices, $ \epsilon_{\nu \alpha \beta}$  being the Levi-Civita tensor. From the above definition, it follows that the divergence of $\langle \mathbf{S} \rangle_{\rm eq}$ is given by
\begin{equation}
\begin{split}
	\nabla\cdot\langle \mathbf{S} \rangle_{\rm eq} &=  
    \lim_{\mathbf{r}' \rightarrow \mathbf{r}}\Big[\nabla \cdot \langle  \mathbf{E} (\mathbf{r}, t) \times \mathbf{H}(\mathbf{r}', t) \rangle_{\rm eq}^{\rm sym} \\
    &+\nabla' \cdot \langle \mathbf{E} (\mathbf{r}, t) \times\mathbf{H}(\mathbf{r}', t) \rangle_{\rm eq}^{\rm sym}\Big],
\end{split}
\end{equation}
which can also be written as
\begin{equation}
\begin{split}
	\nabla\cdot\langle \mathbf{S} \rangle_{\rm eq} &= 
    \lim_{\mathbf{r}' \rightarrow \mathbf{r}}\epsilon_{\nu \alpha \beta}\Big[ \partial_\nu \langle E_\alpha (\mathbf{r}, t) H_\beta(\mathbf{r}', t) \rangle_{\rm eq}^{\rm sym} \\
    &+\partial'_\nu\langle E_\alpha (\mathbf{r}, t) H_\beta(\mathbf{r}', t) \rangle_{\rm eq}^{\rm sym}\Big].
\end{split}
    \label{Eq:Divergence_Poynting}
\end{equation}
We see that $\langle \mathbf{S} \rangle_{\rm eq}$ and its divergence are directly related to the mixed Green's tensors  through the correlation function. In the following sections, we will evaluate these quantities focusing on non-reciprocal systems.

\section{Lorentz reciprocity}

In thermal equilibrium, one would expect the mean Poynting vector to vanish, indicating the absence of energy or heat flux. However, this condition is only guaranteed in electromagnetic systems that preserve time-reversal symmetry. To see this, we employ the Lorentz reciprocity which holds in time-reversal symmetric systems~\cite{Caloz}. 
For the Green functions,
the Lorentz reciprocity is expressed by the well known relations~\cite{GSA,Eckhardt}
\begin{align}
	\mathds{G}^{\rm EH}(\mathbf{r},\mathbf{r}';\omega) &= - {\mathds{G}^{\rm HE}}^t (\mathbf{r}',\mathbf{r};\omega), \label{Eq:LorentzHE} \\
	\mathds{G}^{\rm EE}(\mathbf{r},\mathbf{r}';\omega) &=  {\mathds{G}^{\rm EE}}^t (\mathbf{r}',\mathbf{r};\omega), \label{Eq:LorentzEE}
\end{align}
the superscript $t$ denoting the matrix transpose operation.
An important point to note here is that, when the system is time-reversal symmetric and Lorentz reciprocity holds, then
\begin{equation}
   \psi^{\rm EH}_{\alpha\beta} (\mathbf{r}, \mathbf{r'};\omega) = \Theta(\omega,T) 2 \omega \mu_0
	 (-\ri) \Re \bigl[ \mathds{G}^{\rm EH}_{\alpha\beta}(\mathbf{r},\mathbf{r}';\omega) \bigr]
\end{equation}
is purely imaginary. In consequence, the equal time correlation function in Eq.~(\ref{Eq:Equaltimecorrelation}) vanishes and the mean Poynting vector is zero~\cite{GSA}. Hence, in time-reversal systems in thermal equilibrium there is no heat or energy flux, as is well-known since a long time. 

We want to emphasize that we do not consider the asymmetry in time due to dissipation as a time-reversal symmetry breaking but, following Ref.~\cite{Caloz}, as an expression of thermodynamical macroscopic irreversibility. Therefore, in the following we will make use of the non-reciprocity due to time-reversal symmetry breaking and the assymmetry in time due to dissipation as two different notions.

\section{Non-reciprocal systems\label{sec:non-reciprocal}}

We now turn to the more general case in which the system is not constrained by time-reversal symmetry. In this situation, the Lorentz reciprocity relations [Eqs.~(\ref{Eq:LorentzHE}) and (\ref{Eq:LorentzEE})] no longer apply, and as a result, the mean Poynting vector can be non-zero. This has been shown explicitely for a single nanoparticle~\cite{OTTcircular}, several nanoparticles~\cite{ZhuFan,ZhuEtAl2018,OTTReview}, films and planar cavities~\cite{Silveirinha} using materials with a non-reciprocal optical response. As argued by Silveirinha~\cite{Silveirinha} using an eigenmode expansion for the Green's function, even though the mean Poynting vector is non-zero, its the divergence must vanish in thermal equilibrium:
\begin{equation}
  \nabla\cdot\langle \mathbf{S} \rangle_{\rm eq} = 0.
\label{Eq:Divergence}
\end{equation}
This conclusion also follows from the approach in Ref.\cite{OTTcircular}, which considers a single non-reciprocal spherical nanoparticle, and was explicitly demonstrated in Ref.~\cite{Latella2025} to hold for $N$-particle systems in the dipolar approximation. As already pointed out by Silveirinha~\cite{Silveirinha}, the vanishing divergence of the mean Poynting vector ensures that no heat is produced or absorbed in any part of the system. Consequently, there is no net heat transfer in equilibrium, as required by the second law of thermodynamics.

We now evaluate $\langle \mathbf{S} \rangle_{\rm eq}$ in the vacuum region of an otherwise general non-reciprocal electromagnetic system in thermal equilibrium, and demonstrate that it is divergence-free.
From Eqs.~(\ref{Eq:Psispec}) and (\ref{Eq:Equaltimecorrelation}), we have
\begin{equation}
    \begin{split}
        \langle E_\alpha (\mathbf{r}, t) H_\beta(\mathbf{r}', t) \rangle_{\rm eq}^{\rm sym} &= -\int_{0}^{\infty} \frac{\rd \omega}{2 \pi} \Theta(\omega,T)\\ 
        &\times 2\Re \Bigl[ \ri\omega \mu_0\mathds{G}^{\rm EH}_{\alpha\beta}(\mathbf{r},\mathbf{r}';\omega)\\
        &+ \ri\omega \mu_0{\mathds{G}^{\rm HE}_{\beta\alpha}}(\mathbf{r}',\mathbf{r};\omega) \Bigr].
    \end{split}
\end{equation}
By using Eqs.~(\ref{Eq:GEHEE}) and (\ref{Eq:GHEEE}) in the form
\begin{align}
	\mathds{G}_{\alpha \beta}^{\rm EH}(\mathbf{r,r'};\omega) &=\frac{1}{\ri \omega \mu_0} \epsilon_{\beta \gamma \delta} \partial'_\delta \mathds{G}_{\alpha \gamma}^{\rm EE}(\mathbf{r,r'};\omega),   \\
	\mathds{G}_{\beta \alpha}^{\rm HE}(\mathbf{r',r};\omega) &=\frac{1}{\ri \omega \mu_0} \epsilon_{\beta \gamma \delta} \partial'_\gamma \mathds{G}_{\delta \alpha}^{\rm EE}(\mathbf{r',r};\omega), 
\end{align}
we get
\begin{equation}
    \begin{split}
        \langle E_\alpha (\mathbf{r}, t) H_\beta(\mathbf{r}', t) \rangle_{\rm eq}^{\rm sym} &= \int_{0}^{\infty} \frac{\rd \omega}{2 \pi} \Theta(\omega,T)   \epsilon_{\beta \delta \gamma}\\
        &\times 2\Re \Bigl[  \partial'_\delta \mathds{G}_{\alpha \gamma}^{\rm EE}(\mathbf{r,r'};\omega) \\
        &+  \partial'_\gamma \mathds{G}_{\delta \alpha}^{\rm EE}(\mathbf{r',r};\omega) \Bigr].
    \end{split}
    \label{Eq:Average_fields}
\end{equation}
Therefore, from Eqs.~(\ref{Eq:Pontingvector_components}) and (\ref{Eq:Average_fields}) we obtain the components of the equilibrium mean Poynting vector as
\begin{equation}
\begin{split}
	\langle S_\nu \rangle_{\rm eq} &= \lim_{\mathbf{r}' \rightarrow \mathbf{r}} \int_{0}^{\infty} \frac{\rd \omega}{2 \pi} \Theta(\omega,T)   \epsilon_{\nu \alpha \beta}\epsilon_{\beta \delta \gamma}\\
        &\times 2\Re \Bigl[  \partial'_\delta \mathds{G}_{\alpha \gamma}^{\rm EE}(\mathbf{r,r'};\omega)
        +  \partial'_\gamma \mathds{G}_{\delta \alpha}^{\rm EE}(\mathbf{r',r};\omega) \Bigr],
\end{split}
\end{equation}
which is valid in any non-reciprocal system and will be used below to describe some examples.

The divergence of the equilibrium mean Poynting vector can now be evaluated from Eq.~(\ref{Eq:Divergence_Poynting}) by using  Eq.~(\ref{Eq:Average_fields}), leading to
\begin{equation}
	\nabla\cdot\langle \mathbf{S} \rangle_{\rm eq} = \int_0^\infty \!\!\frac{\rd \omega}{2 \pi} \, \Theta(\omega, T) 2 \Re\bigl[ X(\mathbf{r},\omega) \bigr]
	\label{Eq:divergencePV}
\end{equation}
with
\begin{equation}
\begin{split}
	X(\mathbf{r},\omega)
    & =\lim_{\mathbf{r}' \rightarrow \mathbf{r}}\Big\{
        \nabla \cdot  \mathds{G}^{\rm EE}(\mathbf{r',r}) \cdot \nabla' - \nabla' \cdot \mathds{G}^{\rm EE}(\mathbf{r,r'}) \cdot \nabla \\ 
    & - \Big(\nabla \cdot  \nabla' + {\nabla'}^2\Big) \Tr \big[ \mathds{G}^{\rm EE}(\mathbf{r',r}) - \mathds{G}^{\rm EE}(\mathbf{r,r'}) \big] \\
    & + \nabla' \cdot \big[ \mathds{G}^{\rm EE}(\mathbf{r',r}) - \mathds{G}^{\rm EE}(\mathbf{r,r'}) \big]\cdot \nabla' \Big\}.
\end{split}
    \label{Eq:Divergence-free}
\end{equation}
In the limit $\mathbf{r}'\rightarrow\mathbf{r}$, for which $\nabla'\rightarrow\nabla$, the expressions in the bracket cancel out, so that $X(\mathbf{r},\omega) = 0$. We emphasize that the Lorentz reciprocity relations [Eqs.~(\ref{Eq:LorentzHE}) and (\ref{Eq:LorentzEE})] were not used in our derivation, meaning that Eq.~(\ref{Eq:Divergence-free}) provides a general proof that the divergence of the mean Poynting vector in thermal equilibrium is always zero.

By looking at the structure of $X(\mathbf{r},\omega)$, we can also introduce a sufficient condition for having $\nabla\cdot\langle \mathbf{S} \rangle_{\rm eq} = 0$, given by the two relations
\begin{align}
	\lim_{\mathbf{r}' \rightarrow \mathbf{r}}\big[\nabla \cdot  \mathds{G}^{\rm EE}(\mathbf{r',r}) \cdot \nabla' - \nabla' \cdot \mathds{G}^{\rm EE}(\mathbf{r,r'}) \cdot \nabla\big]&=0, \label{Eq:condition1} \\
	\lim_{\mathbf{r}' \rightarrow \mathbf{r}}\big[ \mathds{G}^{\rm EE}(\mathbf{r',r}) - \mathds{G}^{\rm EE}(\mathbf{r,r'}) \big]&=0. \label{Eq:condition2}
\end{align}
If the Green's function is known, these relations can be used to verify that the mean Poynting vector is divergence-free at equilibrium, thus avoiding the need for an explicit evaluation. 
Interestingly, when the Lorentz reciprocity $ \mathds{G}^{\rm EE}(\mathbf{r,r'};\omega) =  {\mathds{G}^{\rm EE}}^t (\mathbf{r',r};\omega)$ is fulfilled, these two relations automatically hold for any $\mathbf{r}$ and $\mathbf{r'}$, without taking the limit. Hence, in this case, the divergence of the equilibrium mean Poynting vector vanishes, as expected, since the mean Poynting vector itself is exactly zero.

\section{Applications}

In this section, we consider specific examples of non-reciprocal systems that illustrate the general discussion in Sec.~\ref{sec:non-reciprocal}, and where it can be explicitly shown that the equilibrium mean Poynting vector is divergence-free. We also examine with certain detail the properties of the equilibrium Poynting vector above a substrate and around a spherical nanoparticle.

\subsection{Normal mode expansion}

Here we consider a general normal mode expansion of the Green function~\cite{Hecht} 
\begin{equation}
	\mathds{G}^{\rm EE}(\mathbf{r,r'}) = c^2 \sum_\mathbf{k} \frac{\mathbf{u}_\mathbf{k}^*(\mathbf{r}',\omega_\mathbf{k}) \otimes \mathbf{u}_\mathbf{k}(\mathbf{r},\omega_\mathbf{k})}{\omega_\mathbf{k}^2 - \omega^2},
	\label{Eq:Green_normal_mode}
\end{equation}
as can be done in lossless (lossy systems) in which the normal modes (quasinormal modes) are the solutions of the wave equation~\cite{Hecht}
\begin{equation}
	\biggl(\nabla\times\nabla\times - \frac{\omega_\mathbf{k}^2}{c^2} \mathds{1}\biggr)\mathbf{u}_\mathbf{k}(\mathbf{r},\omega_\mathbf{k}) = 0,
\end{equation}
where $\mathbf{k}$ specifies a mode of the radiation field (characterized by wavevector and polarization) and $\omega_\mathbf{k}$ is the frequency of the mode $\mathbf{k}$, $c$ being the speed of light in vacuum.

We want to show that the derived relations (\ref{Eq:condition1}) and (\ref{Eq:condition2}) are fulfilled in this case, so the associated mean Poynting vector built with the Green's function~(\ref{Eq:Green_normal_mode}) is divergence-free at equilibrium. Since the relation~(\ref{Eq:condition2}) is trivially satisfied by construction, we only need to concentrate on the relation~(\ref{Eq:condition1}).

From Eq.~(\ref{Eq:Green_normal_mode}), it is not difficult to see that
\begin{equation}
\nabla' \cdot \mathds{G}^{\rm EE}(\mathbf{r,r'}) \cdot \nabla= 
 c^2 \sum_\mathbf{k} \frac{\big[\nabla' \cdot\mathbf{u}_\mathbf{k}^*(\mathbf{r}',\omega_\mathbf{k})\big] \big[\nabla\cdot\mathbf{u}_\mathbf{k}(\mathbf{r},\omega_\mathbf{k})\big]}{\omega_\mathbf{k}^2 - \omega^2} 
\end{equation}
and
\begin{equation}
\nabla \cdot \mathds{G}^{\rm EE}(\mathbf{r',r}) \cdot \nabla'= 
 c^2 \sum_\mathbf{k} \frac{\big[\nabla \cdot\mathbf{u}_\mathbf{k}^*(\mathbf{r},\omega_\mathbf{k})\big] \big[\nabla'\cdot\mathbf{u}_\mathbf{k}(\mathbf{r}',\omega_\mathbf{k})\big]}{\omega_\mathbf{k}^2 - \omega^2} 
\end{equation}
are the same in the limit $\mathbf{r}' \rightarrow \mathbf{r}$. Hence, the relation~(\ref{Eq:condition1}) holds, as we wanted to show.

\subsection{Planar substrates}\label{Sec:planar}

As another example, let us consider the region above a non-reciprocal planar substrate which occupies the halfspace $z < 0$ and has an interface which coincides with the $x$-$y$ plane. It has been shown that there is a non-zero mean Poynting vector in thermal equilibrium~\cite{Silveirinha}, i.e.\ a persistent flux. Here we prove explicitly by using the Green's function for this system that the divergence of the mean Poynting vector vanishes in thermal equilibrium.

First, the Green's function can be divided into a vacuum part and a scattering part such that
\begin{equation}
  \mathds{G}^{\rm EE} (\mathbf{r},\mathbf{r'};\omega) = \mathds{G}^{\rm EE,vac} (\mathbf{r},\mathbf{r'};\omega) + \mathds{G}^{\rm EE,sc} (\mathbf{r},\mathbf{r'};\omega).
\end{equation}
The vacuum part $\mathds{G}^{\rm EE,vac} (\mathbf{r},\mathbf{r'};\omega)$ is the usual Green's function of free space without substrate which fulfills Lorentz reciprocity. Hence, the equilibrium mean Poynting vector associated to the vacuum part vanishes as well as its divergence. On the other hand, the scattering part $\mathds{G}^{\rm EE,sc} (\mathbf{r},\mathbf{r'};\omega)$ of the total Green's function takes the presence of the substrate into account, and for a non-reciprocal material the equilibrium mean Poynting vector is in general non-zero. Therefore, we focus now on this scattering part which can be expressed as~\cite{OTTdiode}
\begin{equation}
	\mathds{G}^{\rm EE,sc} (\mathbf{r},\mathbf{r'};\omega) = \int \!\! \frac{\rd^2 \kappa}{(2\pi)^2} \re^{\ri(\mathbf{x - x'})\cdot \boldsymbol{\kappa} } \tilde{\mathds{G}}^{\rm sc} (\boldsymbol{\kappa};\omega) 
	\label{Eq:Green_sc_planar}
\end{equation}
with ($z,z' > 0$)
\begin{equation}
   \tilde{\mathds{G}}^{\rm EE,sc} (\boldsymbol{\kappa};\omega) = \frac{\ri \re^{\ri k_z (z + z')}}{2 k_z} \sum_{k,l = {\rm s,p} } r_{kl} \mathbf{a}^+_k \otimes \mathbf{a}^-_l,
\end{equation}
where we have written the position vectors as $\mathbf{r}=(\mathbf{x},z)$ and $\mathbf{r}'=(\mathbf{x}',z')$ with the components parallel to the surface of the substrate given by $\mathbf{x}=(x,y)$ and $\mathbf{x}'=(x',y')$, $\boldsymbol{\kappa} = (k_x,k_y)^t$ is the wavevector parallel to the surface, and $k_z$ is the wavevector component along the surface normal. The coefficients $r_{kl}$ are the Fresnel reflection coefficients for an incident $l$-polarized wave which is reflected into a $k$-polarized wave. Finally, the vectors $\mathbf{a}^\pm_{\rm s}$ and $\mathbf{a}^\pm_{\rm p}$ are the polarization vectors which form a basis together with the wavevector. The explicit form of these vectors and the reflection coefficient can be found in Ref.~\cite{OTTdiode}, for instance.

By using our criterium expressed in relations~(\ref{Eq:condition1}) and (\ref{Eq:condition2}), we now show that the divergence of the mean Poynting vector in thermal equilibrium always vanishes for planar media. With the explicit expression of the scattering part of the Green's function in Eq.~(\ref{Eq:Green_sc_planar}), we need to check the relation~(\ref{Eq:condition1}), while the relation~(\ref{Eq:condition2}) is trivially satisfied. We obtain
\begin{equation}
\begin{split}
	\nabla'\cdot\mathds{G}^{\rm EE,sc}(\mathbf{r,r'})\cdot\nabla &= \int \!\! \frac{\rd^2 \kappa}{(2\pi)^2} \re^{\ri(\mathbf{x - x'})\cdot \boldsymbol{\kappa} } \\ 
	& \quad \begin{pmatrix} -\ri \boldsymbol{\kappa} \\ \ri k_z  \end{pmatrix} \cdot\tilde{\mathds{G}}^{\rm sc} (\boldsymbol{\kappa}) \cdot\begin{pmatrix} \ri \boldsymbol{\kappa} \\ \ri k_z  \end{pmatrix}   
\end{split}
\end{equation}
and
\begin{equation}
\begin{split}
	\nabla\cdot\mathds{G}^{\rm EE,sc}(\mathbf{r',r})\cdot\nabla' &= \int \!\! \frac{\rd^2 \kappa}{(2\pi)^2} \re^{\ri(\mathbf{x' - x})\cdot \boldsymbol{\kappa} } \\ 
	& \quad \begin{pmatrix} -\ri \boldsymbol{\kappa} \\ \ri k_z  \end{pmatrix} \cdot\tilde{\mathds{G}}^{\rm sc} (\boldsymbol{\kappa}) \cdot\begin{pmatrix} \ri \boldsymbol{\kappa} \\ \ri k_z  \end{pmatrix}. 
\end{split}
\end{equation}
Obviously, in the limit $\mathbf{r}' \rightarrow \mathbf{r}$ both equations are the same, so the relation~(\ref{Eq:condition1}) holds. Hence, the divergence of the equilibrium mean Poynting vector above a planar substrate is always zero. 
This conclusion is, of course, also valid for any planar multilayer structure, where an analogous expression for the scattering part of the Green's function can be introduced.

\subsection{Dipolar many-body systems}

In this section, we consider a collection of dipolar objects in thermal equilibrium where the material properties of these objects can be non-reciprocal. This system was considered in the original work on the persistent heat current~\cite{ZhuFan} (see also Refs.~\cite{ZhuEtAl2018,GuoEtAl2019}), the photon thermal Hall effect~\cite{PBAHall,OTTPBAHall}, and the persistent circular heat flux~\cite{OTTcircular}, to mention a few.

The system consists of $N$ particles described as dipoles at positions $\mathbf{r}_j$ in free space, $j=1,\dots,N$. When the system is illuminated with an incident field $\mathbf{E}^\mathrm{inc}(\mathbf{r})$, the total field at a point $\mathbf{r}$ is given by~\cite{SABreview}
\begin{equation}
	\mathbf{E}(\mathbf{r}) = \mathbf{E}^\mathrm{inc}(\mathbf{r})
	+ k_0^2 \sum_{j,l = 1}^N \mathds{G}^{\rm EE, vac}(\mathbf{r,r}_j) \underline{\alpha}^{\rm dr}_{jl} \mathbf{E}^\mathrm{inc}(\mathbf{r}_l),
\end{equation}
where we have introduced the dressed polarizability~\cite{SABreview}
\begin{equation}
	\underline{\alpha}^{\rm dr} = \frac{1}{\varepsilon_0} \mathds{T}^{-1} \mathds{A}
\end{equation}
expressed by the block matrices
\begin{equation}
	\mathds{T}_{ij} = \mathds{1} \delta_{ij} - (1 - \delta_{ij}) k_0^2 \underline{\alpha}_i \mathds{G}^{\rm EE, vac}(\mathbf{r}_i,\mathbf{r}_j) 
\end{equation}
and $\mathds{A}_{ij} = \varepsilon_0 \delta_{ij} \underline{\alpha}_i$
which contain the ``bare'' polarizabilities $\underline{\alpha}_i$ of the dipolar object $i$, $\varepsilon_0$ being the vacuum permittivity. 
If the incident field is produced by an external dipole at position $\mathbf{r}'$, the response of the system accounting for the field at the point $\mathbf{r}$ is characterized by
\begin{equation}
\begin{split}
	\mathds{G}^{\rm EE}(\mathbf{r,r'}) &= \mathds{G}^{\rm EE, vac}(\mathbf{r,r'}) \\
	&+ k_0^2 \sum_{j,l = 1}^N \mathds{G}^{\rm EE, vac}(\mathbf{r,r}_j) \underline{\alpha}^{\rm dr}_{jl} \mathds{G}^{\rm EE, vac}(\mathbf{r}_l,\mathbf{r}'),
\end{split}
    \label{Eq:Green_tensor_many-dipoles}
\end{equation}
which defines the Green's tensor of the system.

With the above $\mathds{G}^{\rm EE}(\mathbf{r,r'})$ and Eqs.~(\ref{Eq:GEHEE}) and (\ref{Eq:GHEEE}), it is straightforward to obtain the Green's functions $\mathds{G}^{\rm HE}(\mathbf{r,r'})$ and $\mathds{G}^{\rm EH}(\mathbf{r,r'})$ which we use to determine the mean Poynting vector, as defined in Eq.~(\ref{Eq:Pontingvector_components}) with Eq.~(\ref{Eq:Psispec}). We obtain 
\begin{equation}
\langle S_\nu \rangle_{\rm eq} = \epsilon_{\nu \alpha \beta} \langle E_\alpha (\mathbf{r}, t) H_\beta(\mathbf{r}, t) \rangle_{\rm eq}^{\rm sym} 
\label{Eq:Poynting_many_body}
\end{equation}
with the correlation function
\begin{equation}
\begin{split}
	\langle E_\alpha (\mathbf{r}, t) H_\beta(\mathbf{r}, t) \rangle_{\rm eq}^{\rm sym} &= \int_0^\infty \!\! \frac{\rd \omega}{2 \pi} \, \Theta(\omega,T) \omega \mu_0 k_0^2  \\
       &\quad \times \sum_{j,l = 1}^N 2 \Im \mathds{Y}^{jl}_{\alpha\beta}(\mathbf{r};\omega) .
\end{split}
\label{Eq:correlation_function_many-dipoles}
\end{equation}
The quantities $\mathds{Y}^{jl}_{\alpha\beta}(\mathbf{r};\omega)$ in Eq.~(\ref{Eq:correlation_function_many-dipoles})
are the components of the matrices~\cite{Ott_2020,Latella2025}
\begin{equation}
	\mathds{Y}^{jl}(\mathbf{r};\omega) = \mathds{G}^{\rm EE, vac}(\mathbf{r}, \mathbf{r}_j) \bigl(\underline{\alpha}^{\rm dr}_{jl} - {\underline{\alpha}^{\rm dr}_{lj}}^t \bigr) \mathds{G}^{\rm EH, vac}(\mathbf{r}_l, \mathbf{r})
	\label{Eq:Yjl}
\end{equation}
in which $\mathds{G}^{\rm EH, vac}$ is related to $\mathds{G}^{\rm EE, vac}$ through Eq.~(\ref{Eq:GEHEE}).
We emphasize that there is a non-zero equilibrium mean Poynting vector if the system is non-reciprocal which here is expressed in terms of the non-reciprocal dressed polarizability, i.e.\ if we have $  \underline{\alpha}^{\rm dr}_{jl} \neq {\underline{\alpha}^{\rm dr}_{lj}}^t$.
This is exactly the condition under which one can have a persistent heat current as originally discussed in a configuration of three spherical magneto-optical nanoparticles in Ref.~\cite{ZhuFan}. 
By using the correlation function introduced above, we now want to demonstrate that the divergence of the equilibrium mean Poynting vector vanishes for any many-body configuration of dipolar objects. This was already shown in Ref.~\cite{Latella2025} and below we present another proof to highlight our general results of Sec.~\ref{sec:non-reciprocal}. 
To see that $\nabla\cdot\langle\mathbf{S}\rangle_{\rm eq} = 0$ in this many-body system, in view of Eqs.~(\ref{Eq:Poynting_many_body}) and (\ref{Eq:correlation_function_many-dipoles}) it suffices to demonstrate that
\begin{equation}
	\epsilon_{\nu \alpha \beta} \partial_\nu \sum_{j,l=1}^N \mathds{Y}^{jl}_{\alpha\beta} = 0
\label{Eq:Y1}
\end{equation}
with the matrices $\mathds{Y}^{jl}$ given by Eq.~(\ref{Eq:Yjl}). 
\begin{widetext}
Employing Eq.~(\ref{Eq:GHEEE}), we find
\begin{equation}
	\frac{\epsilon_{\nu \alpha \beta} \partial_\nu \mathds{Y}^{jl}_{\alpha\beta}}{ \ri \omega \mu_0} = \Tr\biggl[ \mathds{G}^{\rm HE, vac}(\mathbf{r}, \mathbf{r}_j) \bigl(\underline{\alpha}^{\rm dr}_{jl} - {\underline{\alpha}^{\rm dr}_{lj}}^t \bigr) \mathds{G}^{\rm EH, vac}(\mathbf{r}_l, \mathbf{r}) \biggr]. \\
\end{equation}
Using now the reciprocity relations for the vacuum Green's function from Eq.~(\ref{Eq:LorentzHE}) and summing over all $j$ and $l$, we obtain
\begin{equation}
\begin{split}
	\frac{\epsilon_{\nu \alpha \beta}  \partial_\nu \sum_{j,l} \mathds{Y}^{jl}_{\alpha\beta}}{ \ri \omega \mu_0} 
	&= -  \sum_{j,l}\Tr\biggl[ {\mathds{G}^{\rm EH, vac}}^t (\mathbf{r}_j,\mathbf{r}) \underline{\alpha}^{\rm dr}_{jl}  \mathds{G}^{\rm EH, vac}(\mathbf{r}_l, \mathbf{r}) \biggr] 
	+   \sum_{j,l}\Tr\biggl[ {\mathds{G}^{\rm EH, vac}}^t (\mathbf{r}_j,\mathbf{r})  {\underline{\alpha}^{\rm dr}_{lj}}^t \mathds{G}^{\rm EH, vac}(\mathbf{r}_l, \mathbf{r}) \biggr]. 
\end{split}
	\label{Eq:Y2}
\end{equation}
The second term can be rewritten by using the fact that the trace of a matrix and the trace of the transpose of that matrix are the same
\begin{equation}
\begin{split}
	\sum_{j,l}\Tr\biggl[  {\mathds{G}^{\rm EH, vac}}^t  (\mathbf{r}_j,\mathbf{r})  {\underline{\alpha}^{\rm dr}_{lj}}^t \mathds{G}^{\rm EH, vac}(\mathbf{r}_l, \mathbf{r}) \biggr] 
	&=   \sum_{j,l}\Tr\biggl[ {\mathds{G}^{\rm EH, vac}}^t (\mathbf{r}_l,\mathbf{r})  {\underline{\alpha}^{\rm dr}_{lj}} {\mathds{G}^{\rm EH, vac}} (\mathbf{r}_j, \mathbf{r}) \biggr] \\
      &  =   \sum_{j,l}\Tr\biggl[ {\mathds{G}^{\rm EH, vac}}^t (\mathbf{r}_j,\mathbf{r})  {\underline{\alpha}^{\rm dr}_{jl}} {\mathds{G}^{\rm EH, vac}} (\mathbf{r}_l, \mathbf{r}) \biggr], 
\end{split}
\label{Eq:equality}
\end{equation}
where in the second line we have exchanged the dummy indices $j$ and $l$.
By using Eq.~(\ref{Eq:equality}) in Eq.~(\ref{Eq:Y2}) leads to Eq.~(\ref{Eq:Y1}), and therefore the divergence of the equilibrium mean Poynting vector is exactly zero. To see that also in this case the relations~(\ref{Eq:condition1}) and (\ref{Eq:condition2}) are fulfilled, it is enough to consider only the scattering part of the Green's tensor~(\ref{Eq:Green_tensor_many-dipoles}). Then, it is immediate to get
\begin{equation}
\lim_{\mathbf{r}' \rightarrow \mathbf{r}}\big[ 
\nabla\cdot\mathds{G}^{\rm EE, vac}(\mathbf{r}',\mathbf{r}_j) \underline{\alpha}^{\rm dr}_{jl} \mathds{G}^{\rm EE, vac}(\mathbf{r}_l,\mathbf{r})\cdot\nabla' 
-\nabla '\cdot\mathds{G}^{\rm EE, vac}(\mathbf{r},\mathbf{r}_j) \underline{\alpha}^{\rm dr}_{jl} \mathds{G}^{\rm EE, vac}(\mathbf{r}_l,\mathbf{r}') \cdot\nabla
\big]=0,
\end{equation}
so the relation~(\ref{Eq:condition1}) holds. The relation~(\ref{Eq:condition2}) is trivially satisfied as well.
\end{widetext}

\section{Persistent currents}

Our previous analysis shows that, for a many-body configuration of non-reciprocal dipolar objects in free space, there exists a nonzero energy flux with vanishing divergence under global thermal equilibrium.
This means that, even though there is an energy flux expressed by the non-zero Poynting vector, there is no energy or heat produced or lost in any part of the system. To illustrate this situation, let us consider the three nanoparticles shown in Fig.~\ref{Sketch}(a) in thermal equilibrium, as investigated in the work on the persistent heat current in Ref.~\cite{ZhuFan}. We can choose any of the three nanoparticles, label it as particle $i$, and enclose it within a volume $V_i$. The power emitted or absorbed by the particle is then given by the projection of the mean Poynting vector on the surface normal $\mathbf{n}$ integrated along the whole closed surface $\partial V_i$ of that volume, so that 
\begin{equation}
	P_i = \int_{\partial V_i} \!\!\! \rd^2 a\, \mathbf{n} \cdot \langle \mathbf{S} \rangle_{\rm eq} = \int_{V_i} \rd^3 r \, \nabla \cdot \langle \mathbf{S} \rangle_{\rm eq} = 0. 
\end{equation}
Since $\nabla \cdot \langle \mathbf{S} \rangle_{\rm eq}$ vanishes, there is no net heat flux entering or leaving the particle, i.e.\ there is no heat transfer as it should be in thermal equilibrium. However, despite thermal equilibrium, it is possible to define an exchanged power $P_{i \rightarrow j}$ between two distinct particles $i$ and $j$. As discussed in Refs.~\cite{ZhuFan,ZhuEtAl2018,OTTReview}, one finds that $P_{i \rightarrow j} \neq P_{j \rightarrow i}$ for non-reciprocal particles in thermal equilibrium, such that~\cite{ZhuFan,ZhuEtAl2018,OTTReview}
\begin{equation}
	P_{1 \rightarrow 2} = P_{2 \rightarrow 3} = P_{3 \rightarrow 1}
\end{equation}
and
\begin{equation}
	P_{1 \rightarrow 3} = P_{3 \rightarrow 2} = P_{2 \rightarrow 1}.
\end{equation}
Thus, there seems to be a persistent heat current which is connected to the persistent energy flux expressed by the non-zero mean Poynting vector. Because there is no net heat flux, however, one also has~\cite{Latella_PRL2017}
\begin{equation}
	P_i = \sum_{j \neq i} \bigl(P_{i \rightarrow j} - P_{j \rightarrow i} \bigr) = 0.
\end{equation}
Therefore, this naturally raises the question of whether it is possible to measure the persistent heat current or the associated energy flux. 

A possibility for this measurement is suggested in Ref.~\cite{GuoEtAl2019}, where a formal relation between the persistent heat current~\cite{ZhuFan} and the photon thermal Hall effect~\cite{PBAHall} is studied. 
The proposed approach is to consider an out-of-equilibrium scenario realizing the photon thermal Hall effect and measure the heat transfer under small temperature differences as an indirect indication of a persistent heat current. Due to the formal equivalence of $P_{i \rightarrow j}$ in both equilibrium and non-equilibrium conditions, such an interpretation might seem justified. Nevertheless, by definition, the persistent heat current exists only in thermal equilibrium. Therefore, any measurement performed in an out-of-equilibrium setting does not constitute direct or indirect evidence of its existence.
Furthermore, due to the divergence-free nature of the equilibrium mean Poynting vector, one can formally define alternative expressions for $P_{i\rightarrow j}$ in thermal equilibrium, all of which remain valid as long as there is no net transfer of heat.
These $P_{i\rightarrow j}$ do not necessarily have any resemblance to their out-of-equilibrium counterparts, and as a result, no direct connection can be established between the persistent heat current at equilibrium and the out-of-equilibrium heat transfer current.

In addition, one can formally argue that, in an out-of-equilibrium situation, the full mean Poynting vector $\langle \mathbf{S} \rangle$ is a combination of equilibrium and non-equilibrium contributions 
\begin{equation}
	\langle \mathbf{S} \rangle = \langle  \mathbf{S} \rangle_{\rm eq} (T)  + \langle  \mathbf{S} \rangle_{\rm neq} (\Delta T),
\end{equation}
as derived explicitly for dipolar objects in Refs.~\cite{OTTcircular,Latella2025}. Here $T$ is the temperature of the environment and $\Delta T$ is a short-hand notation of all temperature differences $T - T_i$ between the environment and the objects. 
On the one hand, global thermal equilibrium is reached for $\Delta T \rightarrow 0$ and the non-equilibrium part $\langle  \mathbf{S} \rangle_{\rm neq} (\Delta T)$ cancels out. 
On the other hand, the divergence of the equilibrium contribution $\langle  \mathbf{S} \rangle_{\rm eq} (T)$ always vanishes and it is clear that  
\begin{equation}
	\nabla\cdot\langle \mathbf{S} \rangle = \nabla\cdot\langle  \mathbf{S} \rangle_{\rm neq} (\Delta T).
\end{equation}
Hence, any heating of the nanoparticle is strictly associated with the non-equilibrium component of the mean Poynting vector, not with its equilibrium part. As a result, measuring heat transfer under out-of-equilibrium conditions does not amount to measuring the persistent heat current, as the observed transfer is not caused by the persistent energy flux $\langle  \mathbf{S} \rangle_{\rm eq} (T)$. Consequently, the photon thermal Hall effect and the persistent heat current are distinct phenomena, even though they are related. The former can be observed in out-of-equilibrium situations, while the latter is a purely equilibrium effect that cannot be detected through heat transfer measurements.

\section{Conclusion}

We have presented a general proof that the mean Poynting vector is divergence-free in global thermal equilibrium, even in non-reciprocal systems. Additionally, we derived a sufficient condition for this behavior and showed that it is fulfilled by a general normal mode expansion of the Green's function. This condition was further verified explicitly for representative non-reciprocal configurations, such as planar substrates and arbitrary assemblies of particles within the dipolar approximation.

Building on these results, we also discussed the limitations of proposed experimental methods for detecting persistent heat currents. In particular, measuring heat transfer in out-of-equilibrium conditions, such as those realized in the photon thermal Hall effect, does not constitute direct or indirect evidence of the persistent heat current. This is because any heat transfer in such scenarios arises solely from the non-equilibrium component of the mean Poynting vector, not from its equilibrium counterpart. Hence, although the photon thermal Hall effect and the persistent heat current are related, they are fundamentally different: the former is measurable under non-equilibrium conditions, while the latter is an intrinsic equilibrium phenomenon that cannot be observed via standard heat transfer measurements.
Furthermore, the experimental observation of forces induced by non-reciprocal effects in systems in equilibrium contradicts the laws of thermodynamics~\cite{Krueger}, which undermines the feasibility of inferring the presence of persistent energy fluxes through such forces.
In summary, while persistent heat currents and energy fluxes represent a subtle and fascinating aspect of non-reciprocal systems in thermal equilibrium, theoretical proposals for their experimental detection remains an open and challenging problem.
%
%

\begin{acknowledgments}
S.-A.\ Biehs gratefully acknowledges financial support from the Niedersächsische Ministerium für Kultur und Wissenschaft (`DyNano').
I.\ Latella acknowledges financial support from the Ministerio de Ciencia, Innovación y Universidades of the Spanish Government through Grant No. PID2021-126570NBI00 (MICIU/FEDER, UE).
\end{acknowledgments}

\end{document}